\documentclass[reprint,amsmath,amssymb,aps,showpacs,superscriptaddress,prl]{revtex4-1}

\usepackage{graphicx}
\usepackage{bm}
\usepackage{epsfig}
\usepackage{amssymb}
\usepackage{amsfonts}
\usepackage{braket}
\usepackage{color}
\usepackage{epstopdf}
\usepackage{hyperref}
\usepackage{float}
\restylefloat{table}
\usepackage{bibentry}

\usepackage{multirow}
\usepackage[caption=false]{subfig}

\newcommand{\ba}{\begin{eqnarray}}
\newcommand{\ea}{\end{eqnarray}}
\newcommand{\bd}{\begin{displaymath}}

\graphicspath{{figures/}}% Put all figures in this directory.

\begin{document}
\title{Resonating Valence Bond States with Trimer Motifs}
\author{Hyunyong Lee}
\affiliation{Department of Physics, Sungkyunkwan University, Suwon 16419, Korea}
\affiliation{Institute for Solid State Physics, University of Tokyo, Kashiwa, Chiba 277-8581, Japan}
\author{Yun-tak Oh}
\affiliation{Department of Physics, Sungkyunkwan University, Suwon 16419, Korea}
\author{Jung Hoon Han}
\email{hanjh@skku.edu}
\affiliation{Department of Physics, Sungkyunkwan University, Suwon 16419, Korea}
\author{Hosho Katsura}
\email{katsura@phys.s.u-tokyo.ac.jp}
\affiliation{Department of Physics, Graduate School of Science,
The University of Tokyo, Hongo, Bunkyo-ku, Tokyo 113-0033, Japan}
\date{\today}

\begin{abstract}
The trimer resonating valence bond (tRVB) state consisting of an equal-weight superposition of trimer coverings on a square lattice is proposed. A model Hamiltonian of the Rokhsar-Kivelson type for which the tRVB becomes the exact ground state is written. The state is shown to have $9^g$ topological degeneracy on genus $g$ surface and support $\mathbb{Z}_3$ vortex excitations. Correlation functions show exponential behavior with a very short correlation length consistent with the gapped spectrum. The classical problem of the degeneracy of trimer configurations is investigated by the transfer matrix method.
\end{abstract}

\maketitle

Resonating valence bond (RVB) wave functions have been a paradigmatic expression of strongly correlated states of matter, with potential bearings on high-T$_c$ cuprates and two-dimensional spin liquid materials\,\cite{anderson87,anderson04}.
The basic building block of RVB is the spin singlet pair, also known as the dimer, made out of two constituent $S=1/2$ spins.
In the insulating state without holes, every site is covered by one and only one dimer in a particular fashion which we denote as $|{\cal D}\rangle$ - the dimer covering. Quantum fluctuation introduces mixing among various dimer coverings, leading to the RVB wave function as the linear superposition of all possible dimer coverings of the lattice, $|{\rm RVB}\rangle = \sum_{\cal D}  A_{{\cal D}}  |{\cal D} \rangle$, with amplitude $A_{{\cal D}}$. Short-range RVB wave functions of the Rokhsar-Kivelson (RK) type allows dimers over the nearest neighbors only and have been extensively investigated on square\,\cite{rokhsar88}, triangular\,\cite{sondhi01,ivanov04}, hexagonal\,\cite{yao12}, and kagome lattices\,\cite{misguich02}. More recently it became apparent that the tensor network approach is another potent way to probe short-range RVB physics\,\cite{schuch12,poiblanc12,poiblanc13,poiblanc14,hylee16}.

Works on RVB have been overwhelmingly focused on $S=1/2$ spins, motivated at first by its potential relevance to the cuprate physics and later by the discovery of several two-dimensional spin liquid materials all formed out of $S=1/2$ constituent spins\,\cite{balents10}. Discovery of the spin liquid phase in the pnictide family of superconductors in recent years\,\cite{wang15}
has nudged the tide, and now there are active discussions of possible spin liquid phases realized in various frustrated $S=1$ spin models\,\cite{jiang09,jiang12,hylee16}. A pair of spin-1 constituents can form a dimer just as a pair of spin-1/2's do, but there is also a more novel possibility of the trimer, made of three constituent spin-1's forming a spin singlet. Examination of the RVB wave function consisting of trimer motifs, rather than dimers, is the thrust of this work.

The physics of RVB is related to the problem of dimer coverings on planar lattices dating back to 1930s\,\cite{fowler37,chang39} and invigorated by their exact solutions found in 1961 by Kasteleyn\,\cite{kasteleyn61}, and Fisher and Temperley\,\cite{fisher61}. The combinatorial problem of counting the number of coverings can be rephrased as the evaluation of certain partition functions of classical statistical mechanics models. In turn, this partition function can be written as the tensor trace of judiciously chosen site tensors\,\cite{lieb67}. The trimer covering problem is the simplest example of generalization of the heavily studied dimer problem. Our strategy is to re-visit some known results on the classical trimer covering problem\,\cite{van75, kaye77, dhar07, klaus96}, generalize them, and use those results to shed light on the quantum superposition of the trimer coverings which we call the trimer RVB state. We argue, in various ways, that low-energy excitations of the trimer RVB is likely to be described by some sort of $\mathbb{Z}_3$ gauge theory with $9^g$ topological degeneracy on genus $g$ surface.

\begin{figure}
\includegraphics[width=0.5\textwidth]{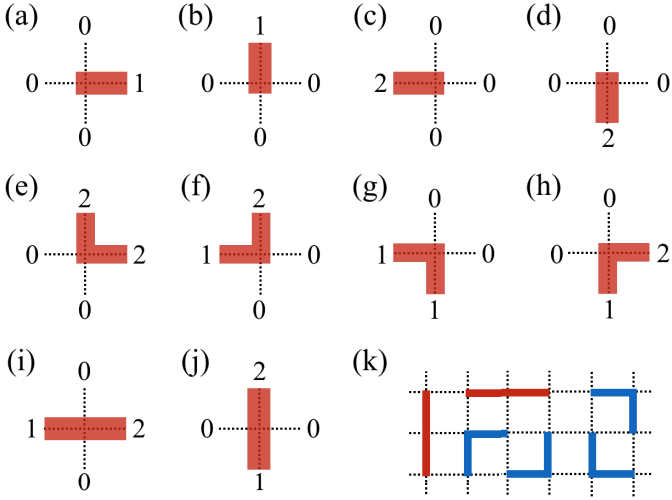}
	\caption{(Color online) (a)-(j) Ten possible configurations of a local tensor to realize the trimer covering. Known trimer coverings consisting of linear and bent types only arise as special limits. The number 0,1 and 2 denote the indices of the site tensor. (k) Red and blue bars represent all possible L-type and B-type trimers, respectively.}
	\label{fig:tm_conf}	
\end{figure}

Different facets of the trimer covering on the two-dimensional square lattice have been investigated for some time\,\cite{van75, kaye77, dhar07, klaus96}. Trimers come in two types: the linear trimer (L-type) which extends either horizontally or vertically over three consecutive lattice sites, or the bent trimer (B-type) extending over a site and its two nearest neighbors as illustrated in Fig. \ref{fig:tm_conf} (k). In this work we consider the square lattice only. Counting the L-trimer covering on a square lattice was done in Ref.\,\cite{dhar07}, and that of the B-type trimer in Ref.\,\cite{klaus96}. Oddly, the counting problem for the mixed case with both types of trimers allowed has never been solved. The mixed trimer problem, to which the L-type and B-type cases belong as special limits, can be formulated using the tensor network language\,\cite{lieb67}.

To this end a site tensor $A_{lrud}$ having the value 1 for the ten configurations shown in Fig.\,\ref{fig:tm_conf} (a)-(j) and zero otherwise is introduced. The indices $l,r,u,d$ refer to left, right, up, and down bonds around each site and take only three possible values 0, 1, and 2. The index $l=0$ implies the absence of trimer in that direction. Two independent values 1 and 2 are introduced for the index designating the presence of trimer bond. As a result, configuration (a) in Fig. \ref{fig:tm_conf} cannot be contracted with (c), nor (b) with (d), to form a dimer. Inspection of the index assignments in Fig. \ref{fig:tm_conf} should convince the readers that only the trimer configurations are allowed under the tensor contraction. L-trimers are formed by contracting (a)-(i)-(c) tensors horizontally, or (b)-(j)-(d) tensors vertically. There are four orientations for B-trimers, each realized by (d)-(e)-(c), (a)-(f)-(d), (a)-(g)-(b), (c)-(h)-(b) contraction, respectively. A trimer model consisting of L-trimers only are achieved as the limit where (e)-(h) configurations are given the value 0.  B-trimer-only model is the limit with (i) and (j) configurations set to zero. Carrying out tensor contraction $Z= \sum_{\{l_i,r_i,u_i,d_i \}} A_{l_1 r_1 u_1 d_1} \delta_{r_1 l_2} A_{l_2 r_2 u_2 d_2} \cdots$ gives the number of trimer coverings.

Efficient calculation of $Z$ proceeds on a cylinder geometry with the periodic boundary condition\,(PBC) imposed along the $x$-direction of length $N_x$ and open ends along the $y$-direction of length $N_y$. The largest eigenvalue $\lambda$ of the row-to-row transfer matrix\,(TM) is used to evaluate $Z \sim \lambda^{N_y}$. Entropy per site is $s= \ln Z/N$, where $N=N_x N_y$ counts the number of lattice sites. We quote below the thermodynamic extrapolation of the entropy for the mixed (both L- and B-type) trimer case along with known results for L-only and B-only types\,\cite{klaus96,dhar07}:
\begin{align}
	s_{\infty} =
	\begin{cases}
    	0.15852 ~ (1.17178) & \quad \text{for L-type}\\
	    0.27693~ (1.31907)  & \quad \text{for B-type}\\
	    0.41194~ (1.50974) & \quad \text{for L- and B-type}\\
  	\end{cases}.
  	\label{eq:tri_entropy}
\end{align}
The number in the parenthesis is $x= e^{s_\infty}$, with which one obtains $Z \sim x^N$.

\begin{figure}[!t]
\includegraphics[width=0.4\textwidth]{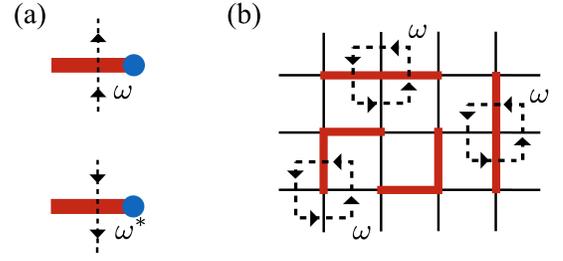}
	\caption{(Color online) (a) Assignment of the weight $\omega=e^{2\pi i/3}$ and its conjugate $\omega^*$ for the passage through the dual lattice with the center of the trimer (blue dot) on the right and left side of the path, respectively. (b) For any elementary loop surrounding a site, the total weight is always $\Gamma = \omega$. For any loop surrounding a single trimer the total weight is $\Gamma = \omega^3 = 1$.}
	\label{fig:string}	
\end{figure}

Interestingly, we have found from analysis of the adjacency graph of the transfer matrix that trimer configurations on the cylinder can be classified into three distinct topological sectors\,\cite{supple}. The winding number characterizing each sector can be defined with a string threading the dual lattice. As depicted in Fig. \ref{fig:string}(a) we assign a direction to the string and give a weight $\omega=e^{2\pi i/3}$ when the center position of the trimer is seen on the right side of the path, and $\omega^*$ if seen on the left side. With this definition, the total weight for the elementary string loop surrounding a site anti-clockwise\,[Fig.\,\ref{fig:string}\,(b)] is always $\omega$, and it can be considered as a locally conserved quantity in the trimer problem. The winding number $\Gamma$ around a single trimer is the product of three factors of $\omega$ as the loop should consist of three consecutive elementary loops, thus giving $\Gamma=\omega^3 = 1$. For non-contractible loops defined on a compact manifold such as the torus, allowed winding numbers are $\Gamma = 1,\,\omega,\,\omega^*$, as readers can easily verify, and cannot be modified by local rearrangements of the string or of the trimers.

\begin{figure}
\includegraphics[width=0.45\textwidth]{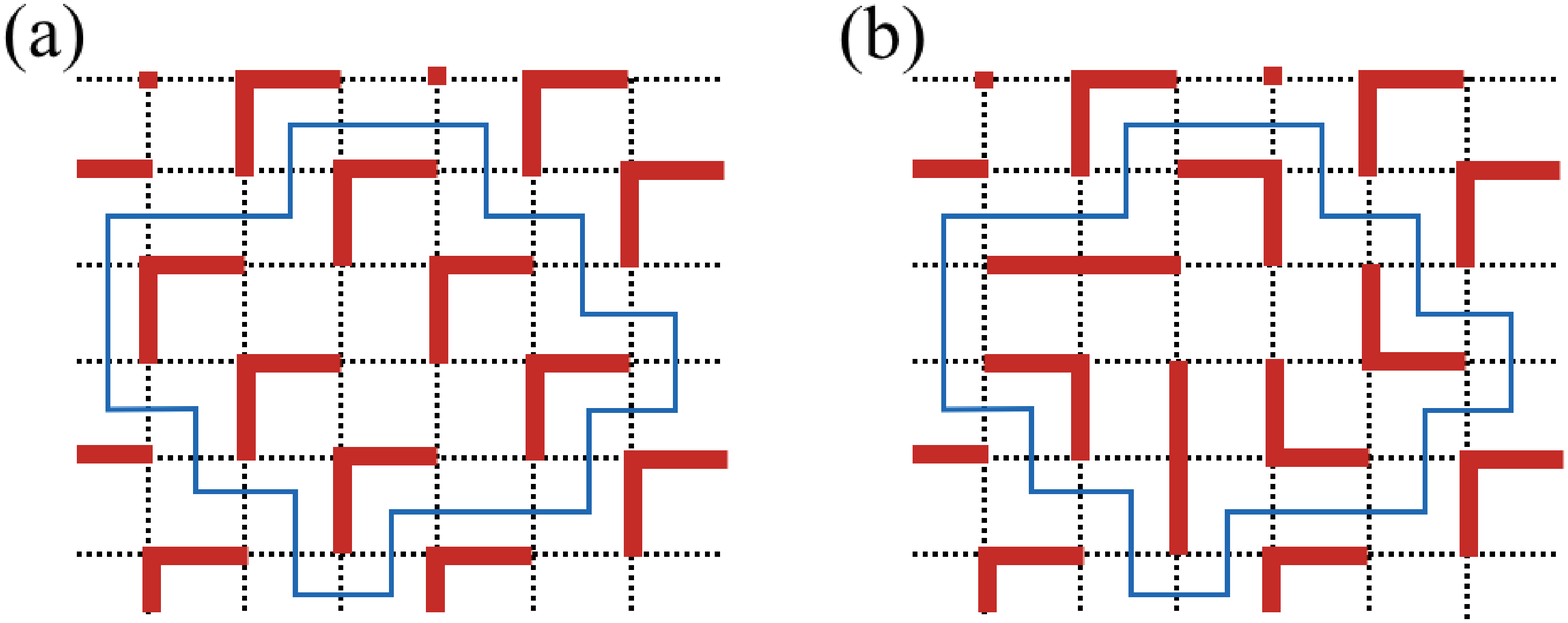}
	\caption{(Color online) (a) Staggered trimer configuration. A non-flippable six-trimer block inside the blue boundary can be connected to (b) a flippable configuration through six-trimer resonance moves.}
	\label{fig:six_trimer_flip}	
\end{figure}

A quantum Hamiltonian with $9^g$ ($g$=genus of the manifold) topological degeneracy may be constructed in analogy with the RK Hamiltonian\,\cite{rokhsar88,sondhi01}:

\begin{align}
	\includegraphics[width=0.48\textwidth]{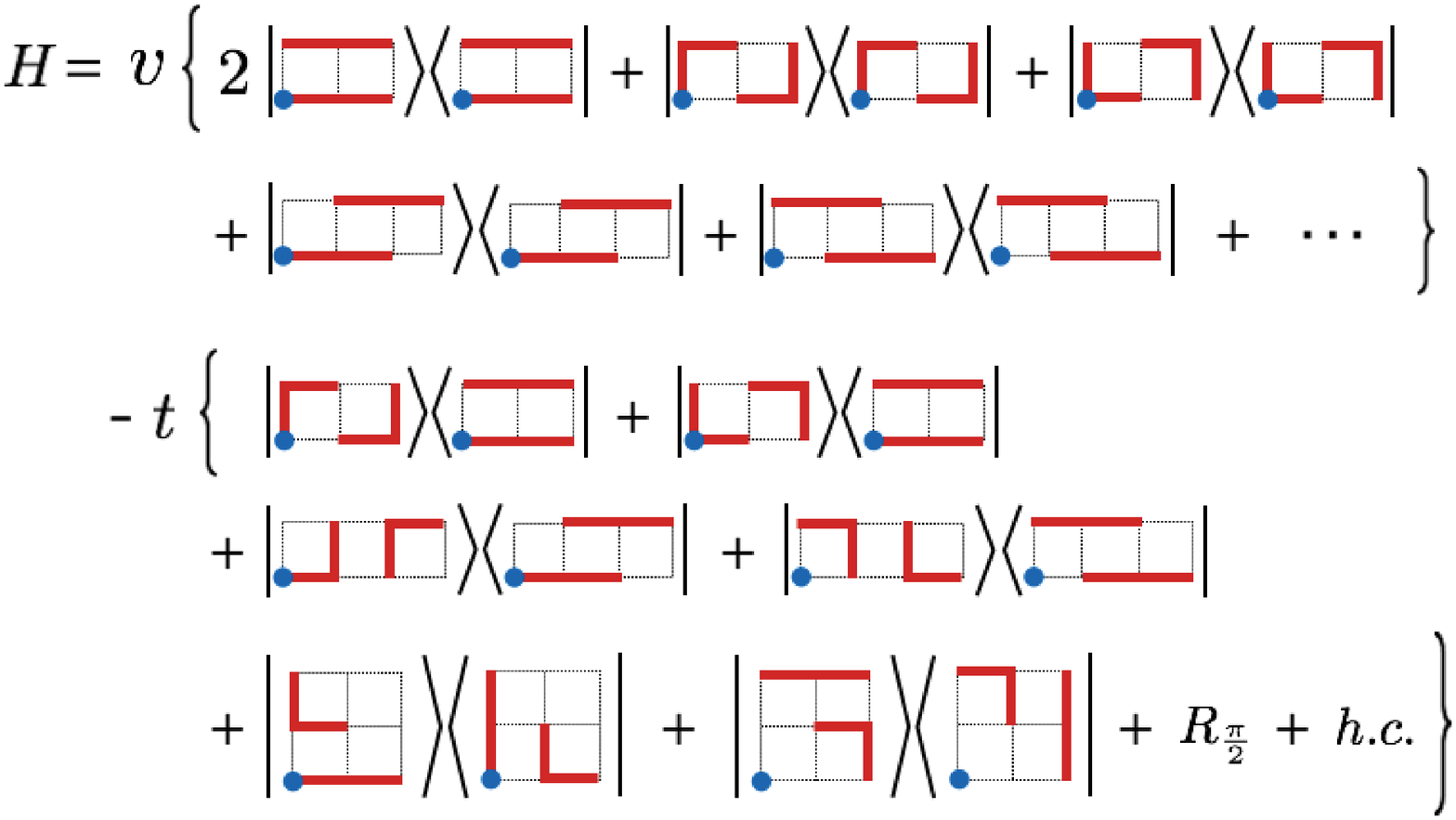} .
	\label{eq:trimer_h}
\end{align}
Summation over all lattice translations of the blue-dot site, as well as the 90 degree rotation ($R_{\frac{\pi}{2}})$ of the displayed terms are assumed. The $v$ and $t$ terms are the potential and resonating pieces, respectively, in analogy with the structure of the RK Hamiltonian for dimers\,\cite{rokhsar88,sondhi01}. In the potential terms, $\cdots$ denotes all other possible diagonal terms \,\cite{supple}. Resonating terms involve only two trimers at once and are not able to alter the topological sector of the initial configuration. As with all dimer models, there are certain ``staggered states" that cannot be reached by applying any number of resonance moves. An example is given in Fig.\,\ref{fig:six_trimer_flip}\,(a). Acting with the Hamiltonian on the staggered state gives 0 irrespective of $t$ and $v$ values. Higher-order moves such as the simultaneous rearrangement of six trimers caged inside the blue contour in Fig.\,\ref{fig:six_trimer_flip}\,(a) and (b) can connect such staggered configurations to flippable ones. Even this six-trimer flip is not able to alter the topological sector.

At $t=v$, the trimer Hamiltonian in Eq.\,\eqref{eq:trimer_h} can be recast as a sum of of projectors that locally project out the linear superposition of flippable configurations\,\cite{rokhsar88,sondhi01,supple}.
Its ground state is the linear superposition of all but the staggered trimer configurations (to be defined shortly), with equal amplitudes,  i.e. the trimer resonating valence bond (tRVB) state: $|\rm{tRVB}\rangle = \sum_{\cal T} |{\cal T} \rangle$. Here $|{\cal T}\rangle$ refers to a trimer covering within a particular topological sector.
For a torus with genus number $g = 1$, we have $ 9^g = 9$ degenerate ground states not connected with each other by resonance moves of the Hamiltonian. Each one is a unique ground state of the trimer RK Hamiltonian at $t=v$, within the subspace that excludes staggered configurations, due to the Perron-Frobenius theorem\,\cite{horn2012matrix}. The staggered states also have zero energy, in apparent degeneracy with the tRVB state $|\rm{tRVB}\rangle$. One can rule out staggered states from the ground state by perturbing away from the RK point to $v = t-\epsilon$ infinitesimally, $\epsilon/t \ll 1$\,\cite{savary16}.
Also, since this perturbation does not mix different topological sectors, we still have $9^g$ independent ground states.

\begin{figure}[h]
	\includegraphics[width=0.5\textwidth]{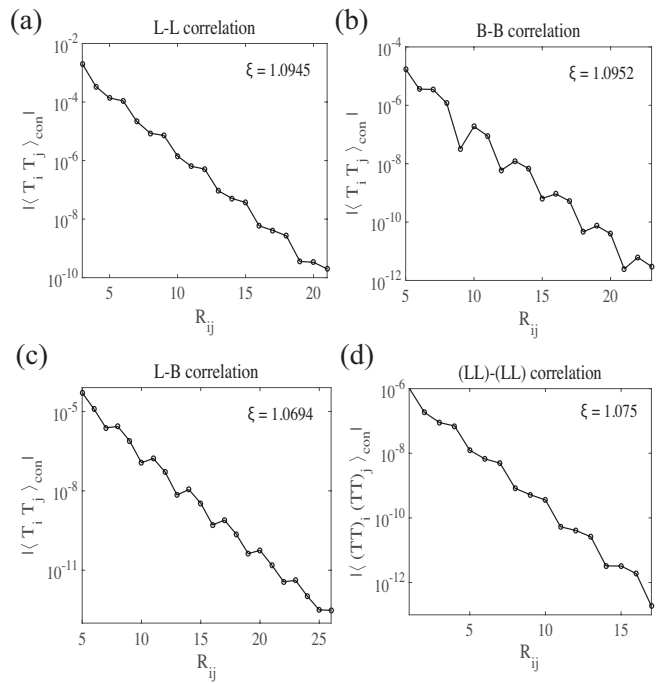}
	\caption{Correlations between (a) L-type, (b) B-type, (c) L- and B-type and (d) (LL)-trimers are measured on $120\times 120$ lattice with the open boundary condition. Estimated correlation length $\xi$ is shown for each plot.}
	\label{fig:trimer_correlations}	
\end{figure}

Connected trimer ($\langle{\rm tRVB}| T_i T_j |{\rm tRVB}\rangle_c$) and trimer-trimer ($\langle{\rm tRVB}| T_i T_{i+\hat{x}} T_{j} T_{j+\hat{x}} |{\rm tRVB}\rangle_c$) correlation functions, where $T_i$ is either a L-type or a B-type trimer projector, are evaluated and presented in Fig.\,\ref{fig:trimer_correlations}. By performing a finite size scaling, we obtained very nicely converged values and therefore the results in Fig.\,\ref{fig:trimer_correlations} are not certainly affected by finite size effect. All functions decay exponentially with very short correlation lengths of order one lattice spacing as observed in the dimer RK wavefunction on the triangular lattice\,\cite{fendley02, ivanov02}. It strongly suggests that the quantum trimer Hamiltonian in Eq.\,\eqref{eq:trimer_h} is gapped at the RK point.

The $9^g$-fold topological degeneracy along with the likely gapped nature of the ground state suggests a $\mathbb{Z}_3$ gauge theory description of the low-energy dynamics for the trimer Hamiltonian. The relevant magnetic excitations (so-called vortex and anti-vortex) will also be of $\mathbb{Z}_3$ character, differentiating a vortex from the anti-vortex\,\cite{schulz12}. (In the $\mathbb{Z}_2$ gauge theory vortex and anti-vortex are the same\,\cite{kitaev06}.) A vortex-anti-vortex pair excitation can be constructed explicitly. Let us consider the same string operator, used previously for defining the winding number, connecting two sites\,($p_1,p_2$) on the dual lattice and define a quantum state

\begin{align}
	|v_1 \bar{v}_2\rangle = \sum_{\mathcal{T}} \omega^{n_r-n_l}|\mathcal{T}\rangle,
	\label{eq:vison_wf}
\end{align}
where $n_{r\,(l)}$ denotes the number of the trimers crossed by the string from the right\,(left) side of its center. This state is orthogonal to the ground state in the thermodynamic limit,

\begin{align}
	\langle v_1 \bar{v}_2| {\rm tRVB}\rangle = \sum_{\mathcal{T}} \omega^{n_l-n_r} \propto 1+\omega+\omega^* = 0 .
\end{align}
The first equality follows from the assumed orthogonality of different trimer configurations $\langle \mathcal{T}' | \mathcal{T}\rangle = \delta_{\mathcal{T}\mathcal{T}'}$. For a sufficiently large sample and a well-separated vortex-anti-vortex pair there should be equal numbers of configurations having $n_l - n_r = 0, 1, 2$ (mod 3), hence the overlap must be zero.
The phase $V_{12}=\omega^{n_l-n_r}$ is topologically identical to the operator creating the $\mathbb{Z}_3$ vortex-anti-vortex pair in the $\mathbb{Z}_3$ gauge theory\,\cite{schulz12}. Therefore, $|{\rm tRVB}\rangle$ and $|v_1 \bar{v}_2\rangle$ can be considered as a vacuum and a single vortex-anti-vortex pair state, respectively. In this sense, we may interpret the nontrivial phase $\omega$ obtained by the elementary loop in Fig.\,\ref{fig:string}\,(b) as a result of braiding between the vortex\,(or anti-vortex) and a $\mathbb{Z}_3$ charge placed at each site\,\cite{schulz12}.

\begin{figure}
	\includegraphics[width=0.5\textwidth]{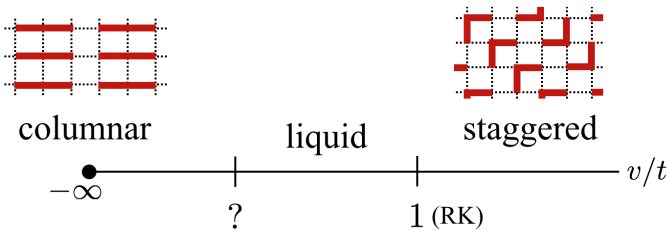}
	\caption{(Color online) Schematic phase diagram of the quantum trimer Hamiltonian.}
	\label{fig:columnar}	
\end{figure}

The $t=v$ RK point defines the first-order phase boundary~\cite{sondhi01}. For $v/t>1$, the ground state is one of the staggered configurations such as Fig.\,\ref{fig:six_trimer_flip}\,(a), defined as states that are annihilated identically by the actions of $t$- and $v$-terms in the trimer Hamiltonian. Any trimer configuration containing a flippable block gains a positive energy $(v-t)n_{fl}$, where $n_{fl}$ is the number of such flippable blocks. At $v/t=-\infty$, the ground state will be chosen to maximize the number of flippable configurations. It is one of the columnar configurations depicted Fig.\,\ref{fig:columnar}, and those are six-fold degenerate at most depending on the boundary condition and the system size. A likely phase diagram of the quantum trimer model is schematically proposed in Fig. \ref{fig:columnar}. An extensive numerical work required to identify the phases and phase boundaries between the trimer RVB state at $v/t=1$ and the columnar phase at $v/t=-\infty$ will be done elsewhere.

The number of staggered configurations in the trimer model grows at least exponentially in the linear dimension. For instance, taking the particular staggered configuration in Fig. \ref{fig:six_trimer_flip} (a) and reversing all the $B$-trimer orientations along one of the north-by-west diagonal line produces another staggered state. That is already $2^{N_d}$ distinct staggered states where $N_d$ counts the number of diagonal rows in the lattice. This is much bigger a number than the linear growth of staggered states in the square-lattice dimer model\,\cite{rokhsar88} or the constant number in the triangular lattice dimer problem\,\cite{sondhi01}. Fortunately, the staggered configurations of the trimer are not allowed in a finite rectangular sample, and therefore the correlation function shown in Fig.\,\ref{fig:trimer_correlations} have been measured in an exact $|\rm{tRVB}\rangle$.

The trimer RVB state is likely to find physical context in the frustrated spin-1 model on a square lattice. A sufficiently strong diagonal antiferrromagnetic exchange interaction $J_2 >0$ in addition to the Heisenberg exchange $J_1$ across the nearest neighbors of the square lattice could favor the formation of a singlet among the three adjacent spins. There is gathering evidence of the spin-liquid phase in the $J_1-J_2$ frustrated spin-1 model. A recent tensor network analysis of possible $S=1$ symmetric spin liquid phases by some of the present authors has found a regime in which the spin-1 RVB states can be realized\,\cite{hylee16}. A careful inspection of the tensor wave function for the spin-1 RVB revealed that the underlying motifs are a mixture of both spin dimers and spin trimers\,\cite{hylee16}.

A final thought on the possible gauge theory description of the trimer dynamics is in order. We know that the constraint of having one and only one dimer per site translates into the Gauss' law constraint on physically allowed states in the gauge description\,\cite{moessner11}. The corresponding constraint for the trimer configuration can be derived with the winding number for an elementary dual-lattice plaquette around each site $\Gamma$ being $\omega$, as discussed earlier. Readers can verify that only the configurations with each site covered by no more than one trimer satisfies this constraint exactly. It seems we have sufficient ingredients regarding the trimer RVB states to declare its candidacy for an example of $\mathbb{Z}_3$ spin liquid. Much more numerical work will be required to prove the claim quantitatively. In the mean time it is rather interesting to speculate that a physical example of $\mathbb{Z}_3$ spin liquid can be written down in terms of a simple generalization of the well-known dimer model.
\\

\acknowledgments{We learned a great deal on the relation of the dimer/trimer covering problem to tensor network from Tomotoshi Nishino's insightful lecture given at ISSP in 2016. HK was supported in part by JSPS KAKENHI Grant No. JP15K17719 and No. JP16H00985. Y.-T.O. was supported by the Global Ph.D. Fellowship Program through the National Research Foundation of Korea (NRF) funded by the Ministry of Education (No. NRF-2014H1A2A1018320).
}

\bibliographystyle{apsrev}
\bibliography{reference}

\clearpage
\onecolumngrid
\begin{center}
\textbf{\large Supplementary Material}
\end{center}

% Fixing numbering of equations/figures: Prefix 'S' and reset counter
\setcounter{equation}{0}
\setcounter{figure}{0}
\setcounter{table}{0}
\setcounter{page}{1}

\begin{center}
\parbox[t][4cm][s]{0.8\textwidth}{
	In this supplementary material, we present the quantum trimer Hamiltonian and the adjacency graphs for the dimer on the square and triangular lattices and the trimer on the square lattice explicitly.}
\end{center}

\twocolumngrid

\section{RK Hamiltonian}
\begin{figure}[!b]	
	\includegraphics[width=0.4\textwidth]{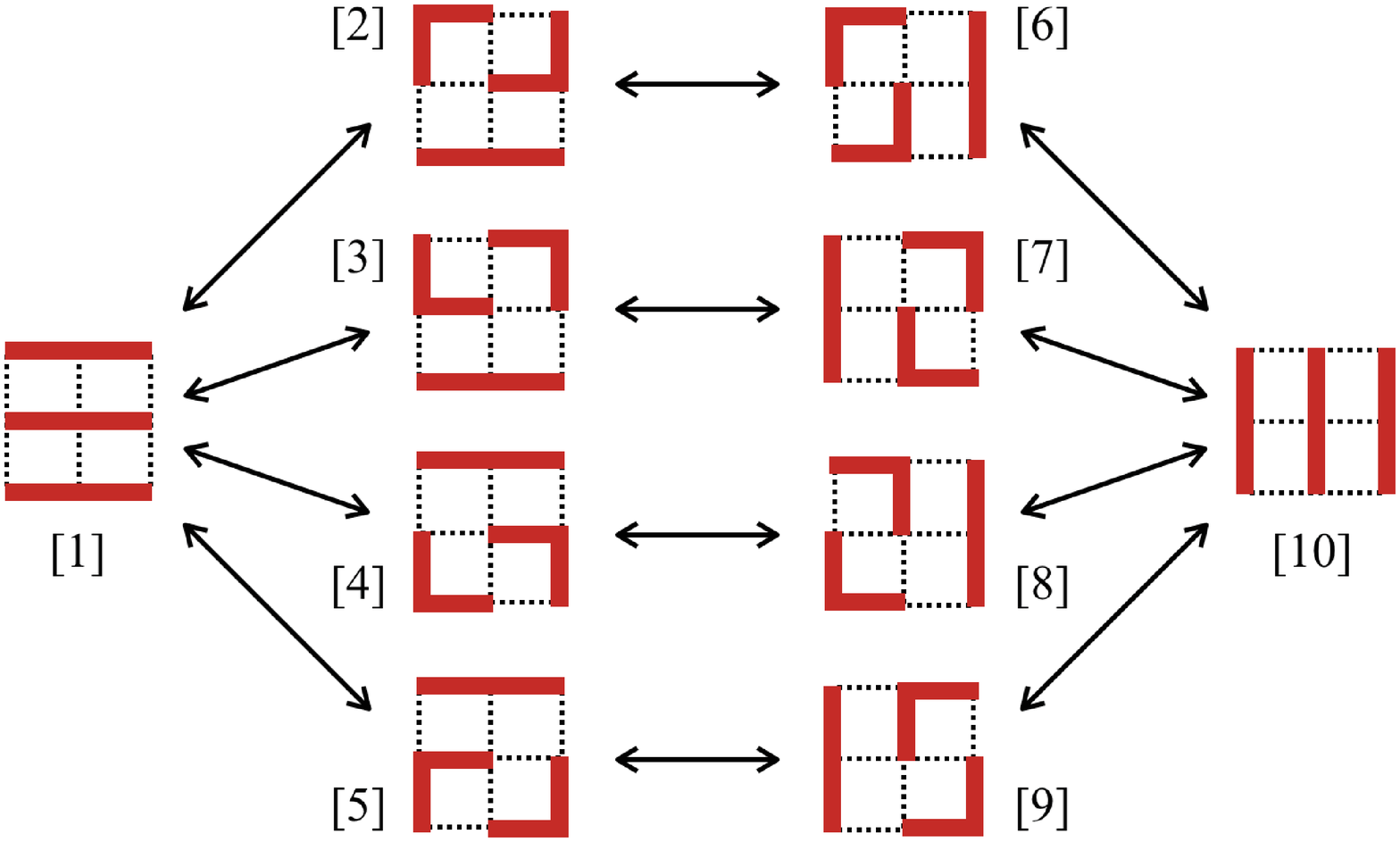}
	\caption{ All possible resonance between the trimer configurations on the $3\times3$ system with the open boundary conditions. Here, the numbers in brackets refer to the basis states. 
% denote the state label. 
Using this basis, one can find the matrix representation of the Hamiltonian\,(see Eq.\,\eqref{eq:h_3by3}).}
	\label{fig:3by3_transition_sm}
\end{figure}

The Rokhsar-Kivelson\,\cite{rokhsar88,sondhi01} type of quantum trimer Hamiltonian may be constructed as follows
\begin{align}
	\includegraphics[width=0.45\textwidth]{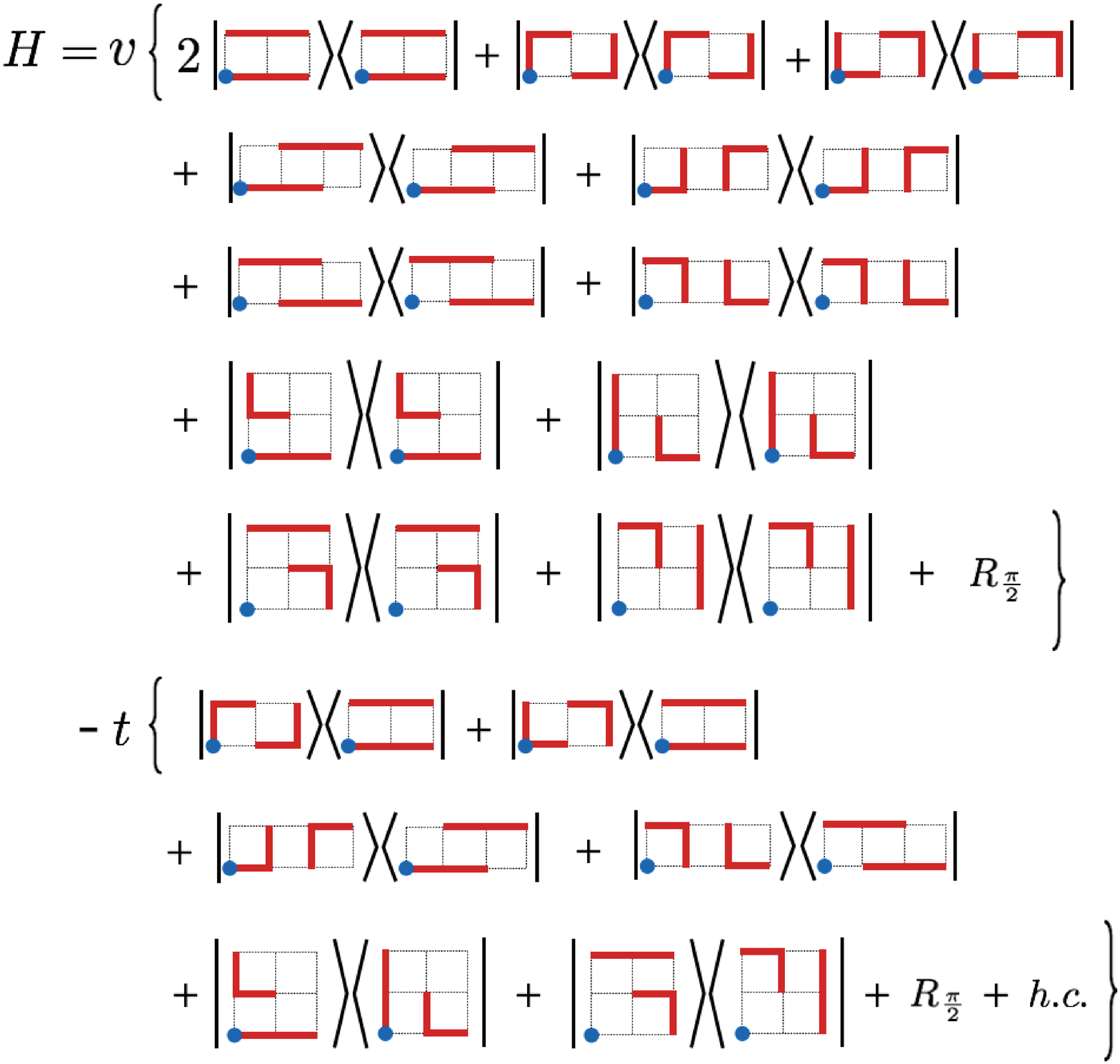},
	\label{eq:trimer_h}
\end{align}
where the summation over all lattice site\,(blue dot) is omitted and $R_{\frac{\pi}{2}}$ in potential and resonating terms stands for $90$ degree rotation of the terms in curly brackets, respectively. At the RK point $t=v=1$, the above Hamiltonian can be recast as 

\begin{align}
	\includegraphics[width=0.45\textwidth]{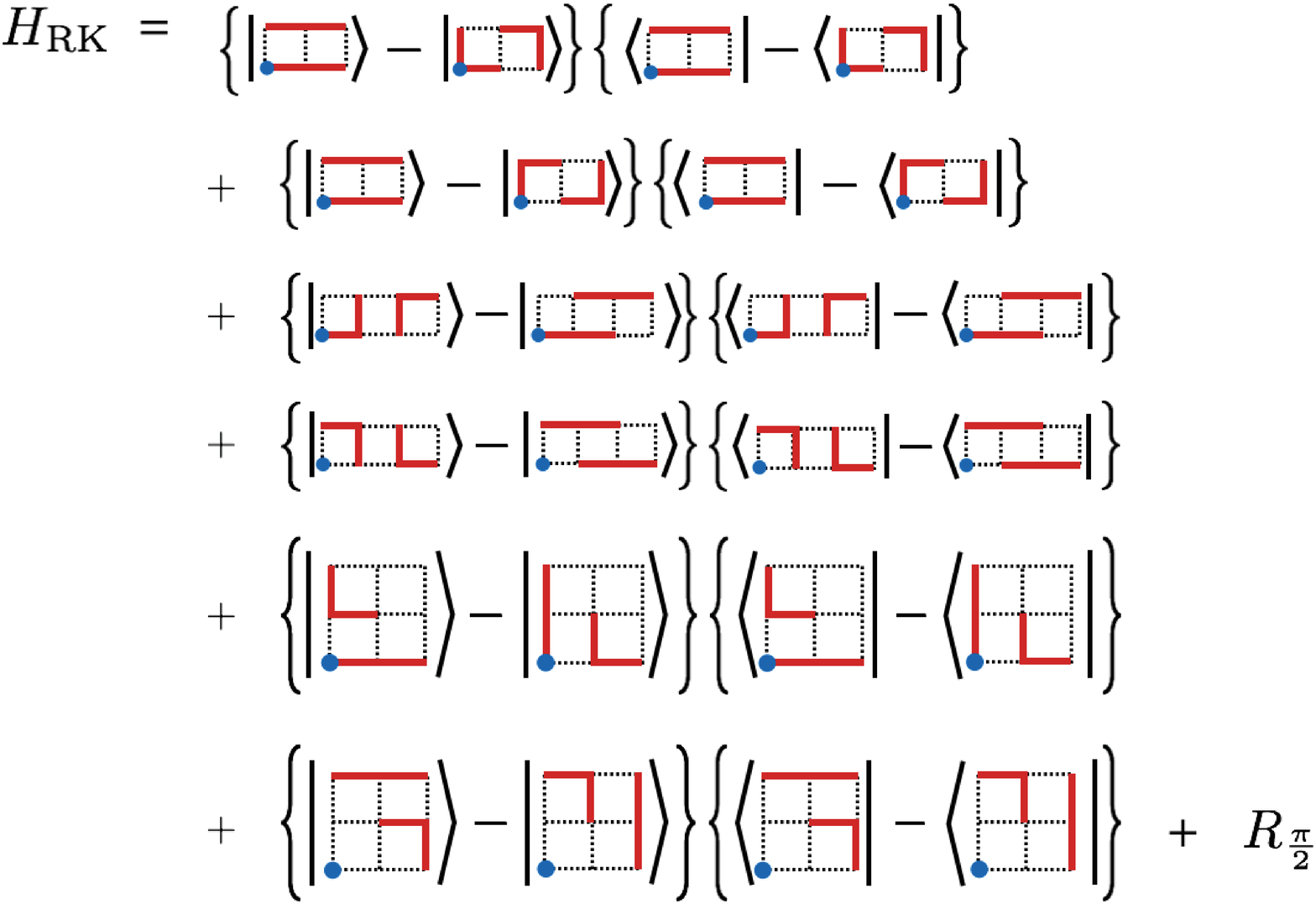},
	\label{eq:trimer_h_RK}
\end{align}
where each term is a projector that locally projects out the linear superposition of flippable configurations. Therefore, the ground state is a linear superposition of all possible\,(excluding the staggered) trimer configurations with the ground state energy $E_G=0$. 

As an example, let us consider a $3\times3$ system with open boundary conditions. There are only 10 configurations, and all possible resonances between them are presented in Fig.\,\ref{fig:3by3_transition_sm}. The number in each bracket denotes the state label, and in this basis, the matrix representation of the Hamiltonian in Eq.\,\eqref{eq:trimer_h_RK} reads as

\begin{align}
	H =
	\begin{bmatrix}
		4&-1&-1&-1&-1&0&0&0&0&0\\
		-1&2&0&0&0&-1&0&0&0&0\\
		-1&0&2&0&0&0&-1&0&0&0\\
		-1&0&0&2&0&0&0&-1&0&0\\
		-1&0&0&0&2&0&0&0&-1&0\\
		0&-1&0&0&0&2&0&0&0&-1\\
		0&0&-1&0&0&0&2&0&0&-1\\
		0&0&0&-1&0&0&0&2&0&-1\\
		0&0&0&0&-1&0&0&0&2&-1\\
		0&0&0&0&0&-1&-1&-1&-1&4
	\end{bmatrix}.	
	\label{eq:h_3by3}
\end{align}
By diagonalizaing above matrix, one can find that the lowest eigenvalue is $\lambda_G=0$ and its corresponding eigenvector is $\vec{G} = (1,1,\cdots,1)^T$. One can also easily check that $\vec{G}$ is an eigenvector of $H$ with eignevalue $0$, because the sum of elements in each row equals $0$. It indicates that the ground state is a linear superposition of all possible configurations with the same weight.

\section{Adjacency graphs}
\begin{figure}[!t]	
	\includegraphics[width=0.5\textwidth]{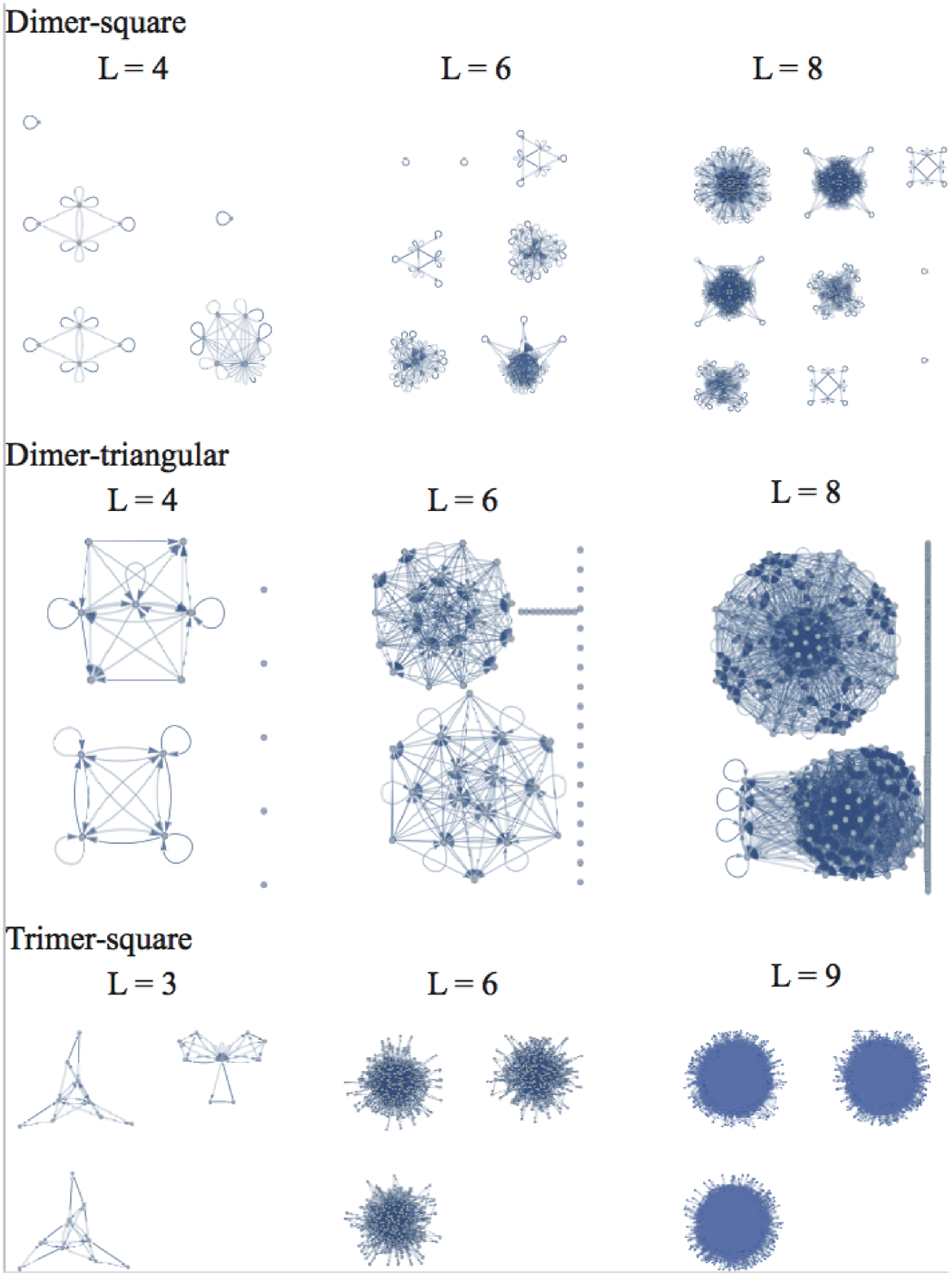}
	\caption{ Adjacency graphs of the transfer matrix for the dimer on the square and triangular lattices and the trimer on the square lattice. Here, $L$ denotes the circumference of the transfer matrix.}
	\label{fig:adj_graphs}
\end{figure}

On the compact spaces such as cylinder and torus in two dimension, the dimer configurations are split into several groups, and the connection between different groups is completely forbidden\,\cite{rokhsar88, sondhi01}. 
The number of groups can be easily checked by constructing a transfer matrix of the partition function and looking into how the adjacency graphs are grouped. To be more specific, we present exemplary adjacency graphs for the dimer on the square and triangular lattices and the trimer on the square latice in Fig.\,\ref{fig:adj_graphs}.
One can define some conserved quantities or winding numbers characterizing each group. These determine a topological character of the RK wave function at RK point. For example, on the square (cylinder) lattice, the number of conserved quantities increases linearly with the linear system size\,(see Fig.\,\ref{fig:adj_graphs}) such that it diverges at thermodynamic limit. Consequently, the RK wave function on the square lattice is considered as a kind of $U(1)$ spin liquid state with gapless excitations\,\cite{rokhsar88}. On the other hand, there are only two conserved quantities on the triangular (cylinder) lattice even at the thermodynamic limit\,(see Fig.\,\ref{fig:adj_graphs}), and the resulting RVB wave function is a gapped $Z_2$ spin liquid state\,\cite{sondhi01}. As R. Moessner and S. L. Sondhi pointed out in Ref.\,\onlinecite{sondhi01}, the number of conserved quantity is related with the gapped or gaplessness nature of excitations. We found that the trimer configurations on the square (cylinder) lattice has three conserved quantities, and they do not depend on the linear system size\,[see Fig.\,\ref{fig:adj_graphs}]. It is similar to the case of dimer on the triangular rather than the square lattice. With the exponential decaying of the correlation functions, the fixed number of conserved quantity strongly suggests that the trimer Hamiltonian is gapped at RK point.

\end{document}